\documentclass{elsart}
\usepackage{graphicx}
\usepackage{epsf}
\usepackage{natbib}
\usepackage{amssymb}

\begin{document}
\begin{frontmatter}
\title{Living in an Irrational Society: Wealth Distribution with Correlations between Risk and Expected Profits}
\author[sfi,cab]{Miguel A. Fuentes}
\ead{fuentesm@cab.cnea.gov.ar}
\author[cab]{M. Kuperman}
\ead{kuperman@cab.cnea.gov.ar}
\author[ufrgs]{J. R. Iglesias *}\corauth[cor]{Corresponding author}
\ead{iglesias@if.ufrgs.br}
\address[sfi]{Santa Fe Institute, 1399 Hyde Park Road, Santa Fe, New Mexico 87501, USA}
\address[cab]{Centro At\'omico Bariloche and Instituto Balseiro,
8400 Bariloche, RN, Argentina}
\address[ufrgs]{Instituto de F\'{\i}sica, UFRGS, C. P. 15051, 91501-970, Porto Alegre, RS, Brazil.}
\date{\today}
\thanks{JRI acknowledges support
from CNPq, Brazil. The authors acknowledge support from CAPES
(Brazil) and SETCYP (Argentina) through the Argentine-Brazilian
Cooperation Agreement BR 085/05. MAF thanks the support of CONICET,
Argentina, and the support and hospitality of Santa Fe Institute,
NM, USA.}
\begin{abstract}
Different models to study the wealth distribution in an artificial
society have considered a transactional dynamics as the driving
force. Those models include a risk aversion factor, but also a
finite probability of favoring the poorer agent in a transaction.
Here we study the case where the partners in the transaction have a
previous knowledge of the winning probability and adjust their risk
aversion taking this information into consideration. The results
indicate that a relatively equalitarian society is obtained when the
agents risk in direct proportion to their winning probabilities.
However, it is the opposite case that delivers wealth distribution
curves and Gini indices closer to empirical data. This indicates
that, at least for this very simple model, either agents have no
knowledge of their winning probabilities, either they exhibit an
``irrational'' behavior risking more than reasonable.
\end{abstract}
\begin{keyword}
econophysics, wealth distribution, Pareto's law \PACS{89.65.Gh \sep
89.75.Fb \sep 05.65.+b \sep 87.23.Ge}
\end{keyword}
\end{frontmatter}

A study of the distribution of the income of workers, companies and
countries was presented, more than a century ago, by Italian
economist Vilfredo Pareto. He investigated data of income for
different European countries and found a power law distribution that
seems to be independent on particular economic condition of each
country. He found~\cite{Pareto} that the distribution of income and
wealth follows a power law behavior where the cumulative probability
$P(w)$ of people whose income is at least $w$ is given by $P(w)
\propto w^{-\alpha}$, where the exponent $\alpha$ is named today
Pareto index. The exponent $\alpha$ for several countries was $1.2
\leq \alpha \leq 1.9$. However, recent data indicate that, even
though Pareto's law provides a good fit to the distribution of the
high range of income, it does not agree with observed data over the
middle and low range of income. For instance, data from
Japan~\cite{souma}, the United States of America and the United
Kingdom~\cite{dragu2000,dragu2001a,dragu2001b} are fitted by a
lognormal or Gibbs distribution with a maximum in the middle range
plus a power law for the highest income. The existence of these two
regimes may be qualitatively justified by stating that in the low
and middle income classes the process of accumulation of wealth is
additive, causing a Gaussian-like distribution, while in the high
income class the wealth grows in a multiplicative way, generating
the power law tail.

Different models of capital exchange among economic agents have been
recently proposed. Most of these models consider an ensemble of
interacting economic agents that exchange a fixed or random amount
of a quantity called ``wealth''. In the model of Dragulescu and
Yakovenko~\cite{dragu2000,review} this parameter is associated with
the amount of money a person has available to exchange, i. e. a kind
of economic ``energy'' that may be exchanged by the agents in a
random way. The resulting wealth distribution is a Gibbs exponential
distribution, as it would be expected. An exponential distribution
as a function of the square of the wealth is also obtained in an
extremal dynamics model where some action is taken, at each time
step, on the poorest agent, trying to improve its economic
state~\cite{PIAV2003,IGPVA2003}. In the case of this last model a
poverty line with finite wealth is also obtained, describing a way
to diminish inequalities in the distribution of
wealth~\cite{SI2004}. In order to try to obtain the power law tail
several methods have been proposed. Keeping the constraint of wealth
conservation a detailed studied proposition is that each agent saves
a fraction - constant or random - of their resources~\cite{review}.
One possible result of those models is condensation, i.e. the
concentration of all the available wealth in just one or a few
agents. To overcome this situation different rules of interaction
have been applied, for example increasing the probability of
favoring the poorer agent in a
transaction~\cite{IGPVA2003,IGVA2004,west1,west2}, or introducing a
cut-off that separates interactions between agents below and above a
threshold~\cite{Das}. Most of these models are able to obtain a
power law regime for the high-income class, but for a limited range
of the parameters, while for the low income, the regime can be
approximately fitted by an exponential or lognormal function.
However, in all those models the risk-aversion (or saving
propensity) of the agents is determined at random with no
correlation with the probability of winning in a given interaction.
Also, possible correlations between wealth and probability of
interaction are not considered.

Here we assume that the agents have some previous knowledge of their
winning probability and they adjust their risk-aversion factor in
correlation with this winning probability. As in previous models we
consider a population of $N=10^5$ interacting agents characterized
by a wealth $w_i$ and a risk aversion factor $\beta_i$. We chose as
initial condition for $w_i$ a uniform distribution between $0$ and
$1000$ arbitrary units. For each agent $i$, the number
$[1-\beta_{i}]$ measures the percentage of wealth he is willing to
risk. At each time step $t$ we select at random the two agents $i$
and $j$ that will exchange resources. Then, we set the quantity to
be exchanged between these two agents as the minimum of the
available resources of both agents, i.e., $dw=\min
[(1-\beta_{i})w_{i}(t);(1-\beta_{j})w_{j}(t)]$. Finally, following
previous works we consider a probability $p \geq 0.5$ of favoring
the poorer of the two partners \cite{IGVA2004,west1},
\begin{equation}
\label{eq:sca} p=\frac{1}{2}+f\times\frac{|w_{i}(t)-w_{j}(t)|}{w_{i}(t)+w_{j}(t)},
\end{equation}
where $f$ is a factor going from $0$ (equal probability for both
agents) to $1/2$ (highest probability of favoring the poorer agent).
Thus, in each interaction the poorer agent has probability $p$ of
earn a quantity $dw$, whereas the richer one has probability $1-p$.
Now we consider that in each transaction both participants know this
probability and adjust their risk-aversion $\beta$ according to the
value of $p$. If the agents are ``rational'' they will risk more
when they have a higher probability of winning so, taking into
account that $p$ varies between $0.5$ and $1$, we first consider
that in each interaction:
\begin{eqnarray}
\label{eq:corr1} \beta_{rich} & = & 2\alpha_r (p-0.5),\nonumber \\
\beta_{poor} & =
 & 2\alpha_p (1-p),
 \label{beta}
\end{eqnarray}
with $\alpha_r$ and $\alpha_p$ ranging from 0 to 1. This
correlation between the risk aversion and $p$ is plotted in Fig.
1, where we change $\alpha_r$ and $\alpha_p$ in order to display
the possible variations of the rich and poor tactics, starting
with a risk-aversion given by $\alpha_r$ = $\alpha_p$ = 1 and then
decreasing the slope from 2 to 0 (so decreasing $\alpha$'s from 1.
to 0.) for the richer agent (Fig.1, left panel) or for the poorer
agent (Fig.1, right panel) up to arriving to a constant risk
aversion equal to zero.
\begin{figure}[ht]
\includegraphics[width=7cm,angle=0]{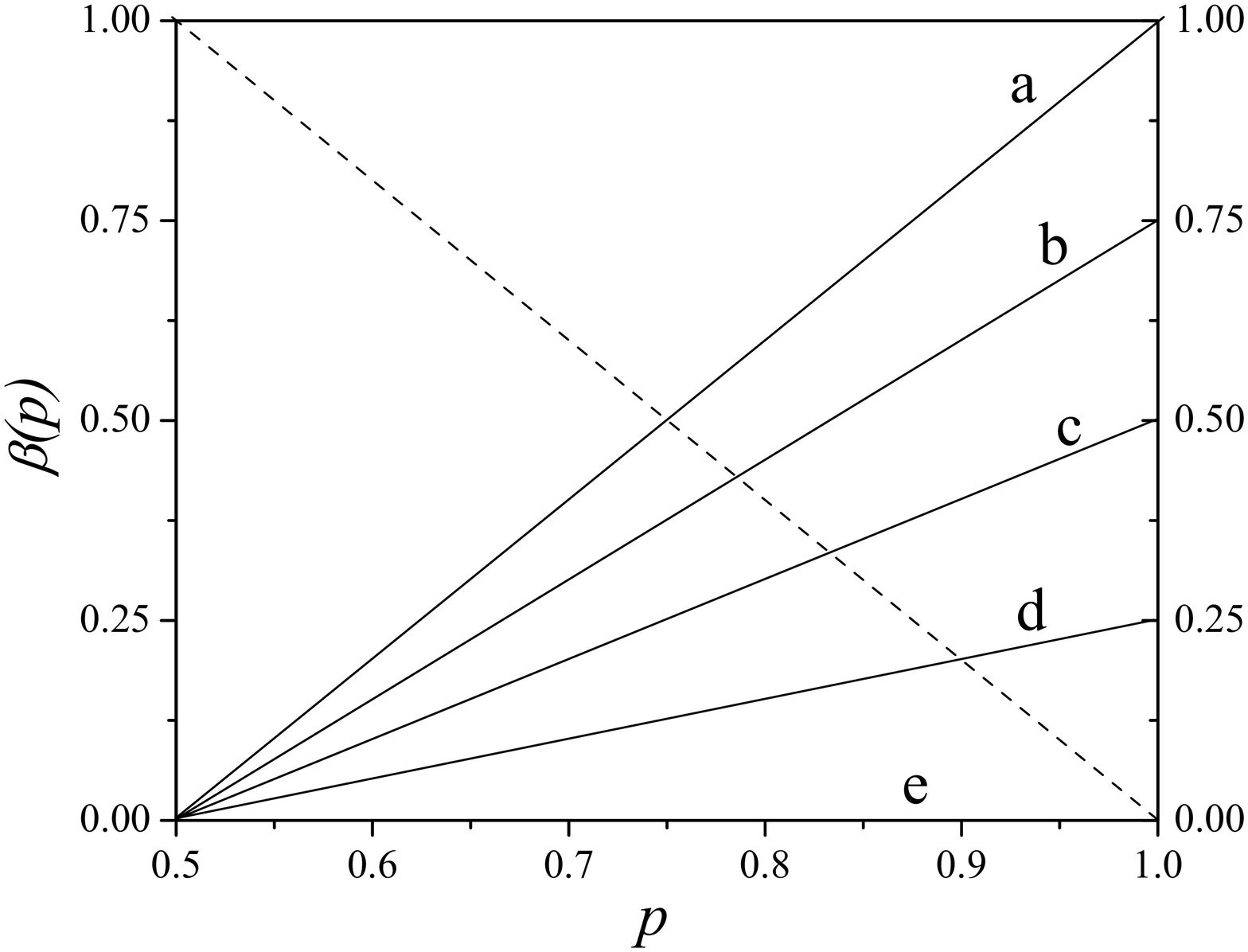}
\includegraphics[width=7cm,angle=0]{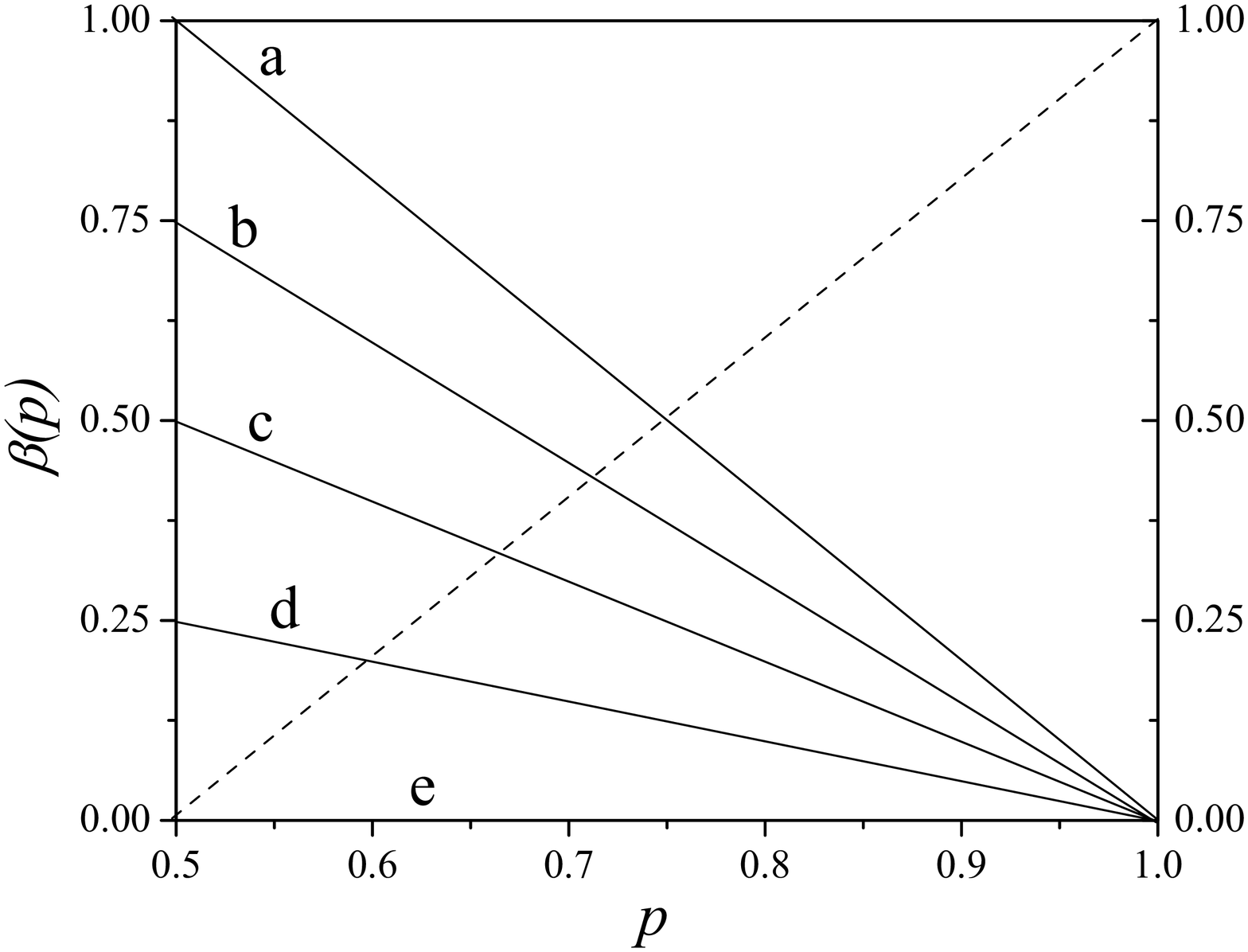}
\caption{{\bf Rational agents}: Risk aversion factor $\beta$ as a
function of $p$, the probability of favoring the poorer of the two
partners. {\it Left panel}: The rich agent, solid lines, change its
behavior going from rational to irrational (lines a to e). Dash line
corresponds to $\beta_{poor}$. {\it Right panel}: Here the behavior
of the poor agent, solid lines, changes, while the dash line
corresponds to $\beta_{rich}$.} \label{fig_1}
\end{figure}

\begin{figure}[ht]
\centering \includegraphics[width=10cm,angle=0]{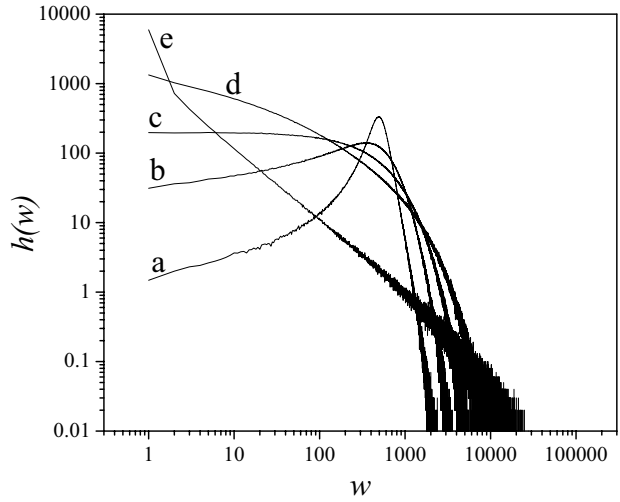}
\caption{Wealth distribution for the case presented in Fig. 1, right
panel, the poor agents change their strategy} \label{fig_2}
\end{figure}

In Fig. 2 we have plotted the wealth distribution corresponding for
changing strategies of the poorer agent. We have not represented the
case when it is the richer agent strategy that changes because we
observe that in this case the wealth distribution is independent of
the changes and is always equal to the curve (a) of Fig.2. Looking
to the (a) curve one observes that a great fraction of the agents
concentrate in a middle class with a wealth very near the average
value and a few agents have wealth bigger than the initial value of
$1000$. This is confirmed by the Gini coefficient of this
distribution that is very low, equal to $0.17$. On the other hand,
the curves (b) to (e) of Fig. 2 correspond to the case when the
risk-aversion of the poorer agents decreases. As it is expected the
inequality increases when the poor agents risk more (and there is
not a change on the rich side). The number of agents with very low
wealth (near $w=1$) increases and for the case where the $\beta$ of
the poor partner is $0$  a power law with an exponent approximately
equal to the unity is obtained. The Gini coefficients also increase
as the risk-aversion of the poor partners decreases as shown in
Fig.~\ref{gini} (open circles), where one can perceive that the Gini
coefficients vary almost linearly from $0.17$ to $0.85$.

A different behavior is obtained when the agents behave
irrationally. Let's modify the behaviors described by equation
(2), that is, the agents risk more when they have a lower chance
to win.
\begin{figure}[ht]
\includegraphics[width=7cm,angle=0]{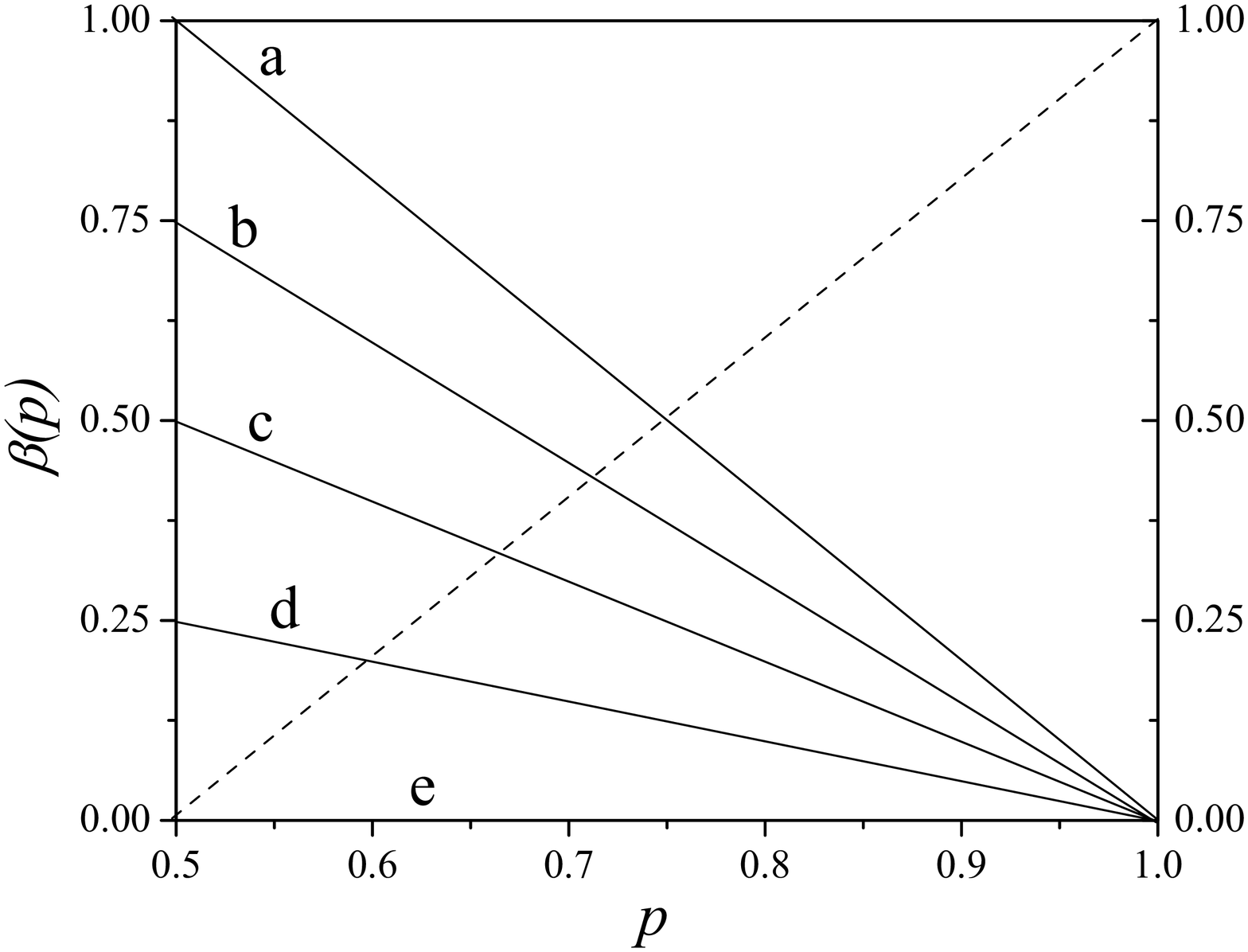}
\includegraphics[width=7cm,angle=0]{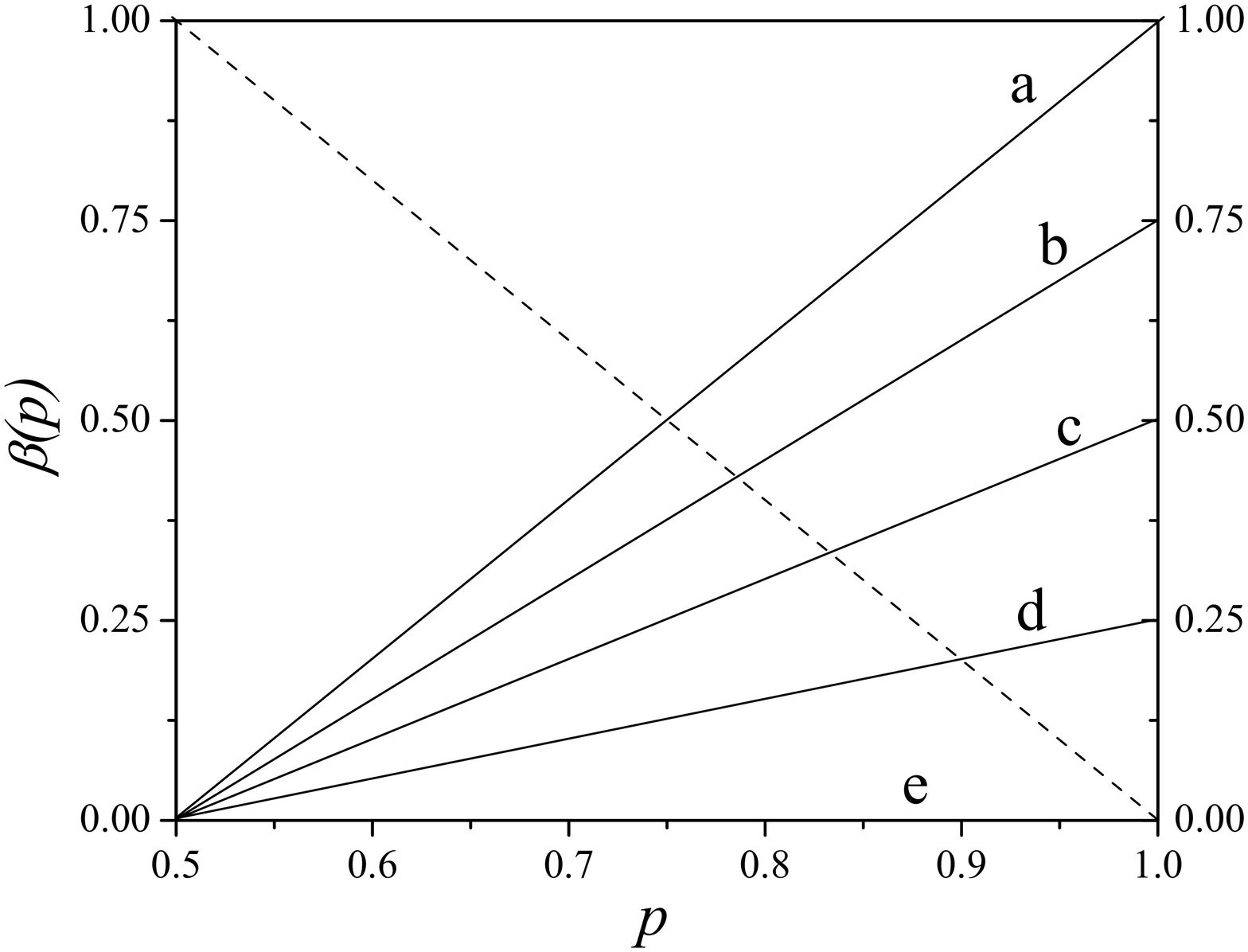}
\caption{{\bf Irrational agents}: Risk aversion factor $\beta$ as a
function of $p$, the probability of favoring the poorer of the two
partners. {\it Left panel}: The rich agent, solid lines, changes its
behavior (lines a to e). Dash line corresponds to $\beta_{poor}$.
{\it Right panel}: The poor agent changes its behavior, solid lines.
Dash line corresponds to $\beta_{rich}$.} \label{fig_3}
\end{figure}
\begin{figure}[ht]
\centering \includegraphics[width=10cm,angle=0]{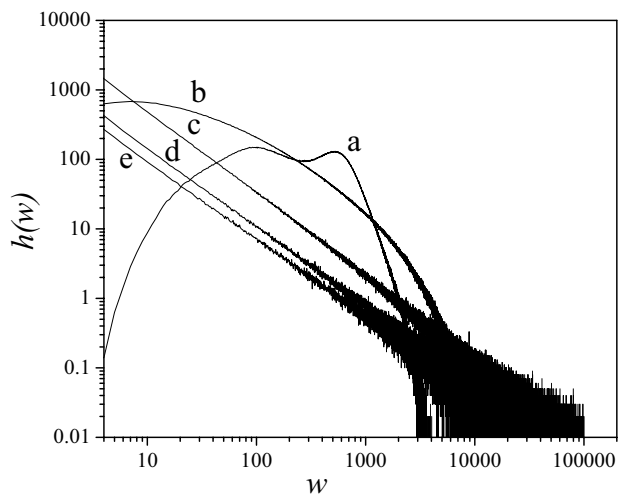}
\caption{Wealth distribution for the case presented in Fig. 3, left
panel, the rich agents change their strategy.} \label{fig_newpot}
\end{figure}
This situation is represented in Fig. 3. If both of them exhibit
an ``irrational behavior'' the effects are mutually neutralized
and the distribution of wealth exhibited in the curve Fig. 4(a)
has a relatively low Gini coefficient, $ \simeq 0.37$. However, if
there is a change in the strategy of the richer agents the effects
are catastrophic for the poorer partners. This result is shown in
curves (b) to (e) of Fig. 4. We can see that the inequality
increases very fast, the wealth distribution approaches to a power
law with an exponent approximately equal to 1.125 and the Gini
coefficients (triangles in Fig. 8) go up to values very near $1.$,
i.e. perfect inequality. On the other hand if the poor agents
change their strategy there are some minor changes in the wealth
distribution, so we have not plotted it. The Gini coefficients are
between $0.35$ and $0.4$ (See Fig. 8, squares) with the exception
of the last point that corresponds to a zero risk-aversion for the
poor partner.

\begin{figure}[ht]
\centering \includegraphics[width=10cm,angle=0]{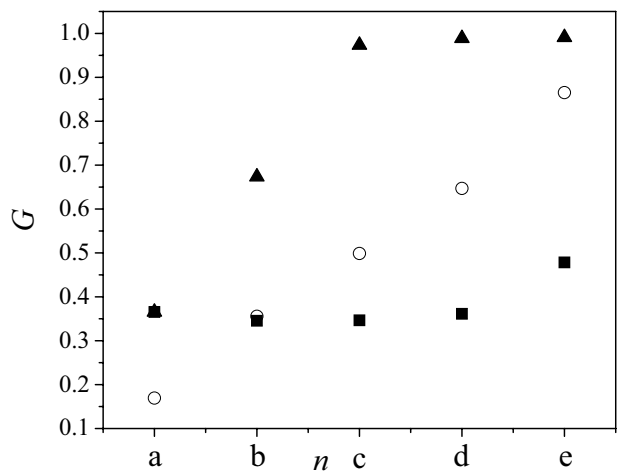}
\caption{Gini coefficients for rational agents (open circles), and
irrational agents when the richer partner changes its
strategy(triangles) or the poorer partner changes its strategy
(squares)} \label{gini}
\end{figure}

The results are summarized in Fig.~\ref{gini} where we have plotted
the Gini coefficient in the different situations discussed above.
The open circles correspond to the rational behavior when the
strategy of the poorer partners changes from the rational to
decreasing values of $\beta$ (so, increasing the risk). One can
observe an almost linear increase of the inequality. That could mean
that when poor agents try to improve their situation risking more,
either because of lack of information, or by trying to improve their
fortune by betting in high risk speculation, the results are in the
opposite direction increasing inequality. On the other hand, in the
case of irrational behavior, when there is a change in the strategy
of the richer partner (triangles), the Gini coefficient increases
very fast up to very high values (near $1$), while if the poorer
partner changes its strategy (squares) there are just minor changes
in the inequality. In any case if one compares our results with Gini
coefficient values for real societies, the values obtained when the
rich partner acts ``rationally'' and the poor partner acts
``irrationally'' -- risking more than reasonable -- are closer to
empirical data, indicating that either the hypothesis of a previous
knowledge of the winning probability is wrong for the poorer
partners, either that the poor agents are in a so bad condition that
they prefer to risk even in the case of a relatively low winning
probability.


\begin{thebibliography}{99}

\bibitem{Pareto} Pareto V (1897), Cours d'Economie Politique, Vol. 2, F. Pichou, Lausanne

\bibitem{souma} Aoyama H, Souma W and Fujiwara Y (2003), Growth and fluctuations of personal
and company's income, Physica A: Statistical Mechanics and its
Applications 324:352--358

\bibitem{dragu2000} Dragulescu A and Yakovenko VM (2000) Statistical Mechanics of Money, The European J. of
Physics B 17:723--729

\bibitem{dragu2001a} Dragulescu A and Yakovenko VM (2001) Evidence for the exponential distribution of income in the USA,
The European J. of Physics B 20:585--589

\bibitem{dragu2001b} Dragulescu A and Yakovenko VM (2001) Exponential and power-law probability distributions of
wealth and income in the United Kingdom and the United States,
Physica A: Statistical and Theoretical Physics 299:213--221

\bibitem{review} See, for instance, ``Econophysics of Wealth
Distribution'' (Springer-Verlag Italia, 2005), ChatterjeeA ,
Yarlagadda S and Chakrabarti BK, eds.

\bibitem{PIAV2003}  Pianegonda S, Iglesias JR, Abramson G and
Vega JL (2003) Wealth redistribution with conservative exchanges
Physica A: Statistical and Theoretical Physics 322:667--675

\bibitem{IGPVA2003} Iglesias JR, Gon\c{c}alves S, Pianegonda S, Vega JL and Abramson G (2003)
Wealth redistribution in our small world, Physica A: Statistical and
Theoretical Physics 327:12--17

\bibitem{SI2004} Pianegonda S and
Iglesias JR (2004) Inequalities of wealth distribution in a
conservative economy , Physica A: Statistical and Theoretical
Physics 42:193--199

\bibitem{IGVA2004} Iglesias JR, Gon\c{c}alves S, Abramson G and Vega JL (2004)
Correlation between risk aversion and wealth distribution, Physica
A: Statistical and Theoretical Physics 342:186--192

\bibitem{west1} Scafetta N, Picozzi S and West BJ (2002)
Pareto's law: a model of human sharing and creativity
cond-mat/0209373v1

\bibitem{west2} Scafetta N, West BJ and Picozzi S (2003)
cond-mat/0209373v1(2002) and A Trade-Investment Model for
Distribution of Wealth cond-mat/0306579v2.

\bibitem{Das} Das A and Yarlagadda S (2005) An analytic treatment of the Gibbs-Pareto behavior in wealth
distribution cond-mat 0409329v1 and to be published in Physica A.

\end{thebibliography}
\end{document}